# Efficient decoy-state quantum key distribution with quantified security

M. Lucamarini,[1,2] K. A. Patel,[1,3] J. F. Dynes,[1,2] B. Fröhlich,[1,2] A. W. Sharpe,[1] A. R. Dixon,[2] Z. L. Yuan,[1,2] R. V. Penty,[3] and A. J. Shields[1,2]

[1]*Toshiba Research Europe Ltd, 208 Cambridge Science Park, Cambridge CB4 0GZ, United Kingdom*
[2]*Corporate Research & Development Center, 1 Komukai-Toshiba-Cho, Saiwai-ku, Kawasaki 212-8582, Japan*
[3]*Cambridge University Engineering Department, 9 JJ Thomson Ave, Cambridge, CB3 0FA, United Kingdom*

**Abstract:** We analyse the finite-size security of the efficient Bennett-Brassard 1984 protocol implemented with decoy states and apply the results to a gigahertz-clocked quantum key distribution system. Despite the enhanced security level, the obtained secure key rates are the highest reported so far at all fibre distances.

**1. Introduction**

In recent years, the possibility to use quantum key distribution (QKD) [1-5] to securely transmit cryptographic keys to remote users has received increasing attention. Consequently, QKD has rapidly grown from early stage experiments to sophisticated demonstrations, suitable for real-world applications [6-10].

To keep pace with the fast progress, high efficiency protocols are demanded. At the same time, it is compelling to demonstrate the security of any such protocol not only in an ideal scenario but also, and more importantly, in the real experimental situation. For example, most of the existing QKD security proofs assume that an infinite dataset is available to the experimenters, the so-called "asymptotic scenario". This leads to overestimate a protocol's security level because the QKD parameters are assumed to be determinable with infinite precision. On the contrary, in a real scenario, the dataset is finite and the security-related quantities are subjected to statistical fluctuations in the sample. Hence, the security level of a real system is also finite and needs to be precisely quantified.

This problem has been theoretically addressed in a few security proofs [11-18], some of which also guarantee composable security [19, 20]. Most of them consider a QKD setup running with the BB84 protocol [2] and making use of an ideal single-photon source. However, in practice, this is not the most convenient choice. Firstly, it is advantageous to use an efficient version of the BB84 protocol [21] rather than the standard one, to increase the key rate of the system. Secondly, an attenuated laser combined with the decoy state technique [22-25] is by far more effective than a single-photon source. Therefore, it would be beneficial to study the security of this more profitable QKD configuration.

To realise that, we purposely designed a protocol, termed "T12", which removes the aforementioned idealisations and encompasses all the features that can possibly increase the practicality of a QKD protocol. Among these, the efficient selection of the basis and that of the light intensity, plus others that will follow from the protocol's security proof. In turn, the security analysis is facilitated by the definition of a protocol, because all the steps leading to the final secure key rate are reunited under a unique description.

For the analysis, we build on previous finite-size security proofs. Among them, the one described in [16, 17, 26] entails several advantages (see also the recent contribution published in [27] and the theoretical studies in [28, 29]). Firstly, it guarantees universally composable security [19, 20], meaning that the key remains secure regardless of the application it is used for, except for some small failure probability ε. Secondly, it includes the squashing model [30, 31] and so the possibility to use threshold detectors. Finally, it is not necessary to perform either a random permutation of the users' strings before the classical post-processing stage [32] or the encryption of error correction information [33].

However, one severe drawback of the aforementioned proof is that when decoy states are taken into account, the key rate remains positive only for very large sample sizes. In fact, in [26] a numerical simulation shows that the data sample must contain at least $10^6$ bits in order to provide a positive key rate. This minimum sample size becomes considerably worse, in fact more than 16 times larger, if the same simulation is run with experimental parameters similar to the ones presented in this work. Moreover, even for a reasonably large sample of $10^8$ bits, the secure key rate is reduced to half its asymptotic value.

Here, we show that the main cause of these problems is the adopted parameter estimation (PE) procedure. Therefore, we review the security proof of [26] and endow it with a more efficient PE, based on numerical optimisation. This improves the results dramatically. We use the new estimates in the security proof to quantify the secure key rate of a gigahertz-clocked QKD system and obtain record rates at all optical fibre distances tested. The approach shows high resistance against small size effects. Using the experimental parameters obtained on a 50 Km optical fibre, the finite-size rate reaches 85% its asymptotic value for a data sample of $10^8$ bits and remains positive for sample sizes as small as $1.4 \times 10^5$ bits.

## 2. Protocol

We start from a description of the T12 protocol. We assume that the transmitter (Alice) has a phase-randomised source of coherent states [34]. This makes the source statistically equal to a Poissonian distribution of number states such that, when the average photon number, or simply the intensity, from the light source is $\mu$, the probability to send a k-photon pulse is Poissonian: $e^{-\mu}\mu^k/k!$. The light pulses are modulated both in intensity and in another degree of freedom, which is used to encode the quantum information. It can be, e.g., the polarisation, or the relative phase from an asymmetric Mach-Zehnder interferometer, like in our setup (see Fig. 1 in Section 5 for a description). For the intensity, Alice randomly chooses among three possible values [25], which we denote with u (signal), v (decoy1) and w (decoy2). It is convenient to introduce a specific intensity label $\mu_j$, with j={0,1,2}, so that $\mu_0=u$, $\mu_1=v$ and $\mu_2=w$. Then the symbol $\mu$ indicates a generic intensity value, while the symbol $\mu_j$ is an intensity index taking on the three specific intensity values u, v and w. The values $\mu_j$ are selected with probabilities $p_{\mu_j}=(p_u,p_v,p_w)$, and usually $p_u \gg p_v > p_w$. For the encoding, Alice randomly selects one of four possible states, as in the standard BB84 protocol [2], indicated as $|0_Z\rangle$, $|1_Z\rangle$ (Z basis) and $|0_X\rangle=(|0_Z\rangle+|1_Z\rangle)/\sqrt{2}$, $|1_X\rangle=(|0_Z\rangle-|1_Z\rangle)/\sqrt{2}$ (X basis). The bases Z and X are selected with probabilities $p_Z \geq 1/2$ and $p_X=1-p_Z$. According to this convention, Z is the majority basis, i.e., the one selected most often, and X the minority basis. When $p_Z > p_X$, there is an increase of efficiency with respect to the standard BB84 protocol, in which $p_X=p_Z$. It is convenient, for the following discussion, to introduce also a basis index

Intensities and states are chosen independently by Alice, so that it is possible to pair any state with any different intensity. This allows for a simpler implementation and prevents accidental correlations between the intensity and the information encoding. In addition, it is possible to distill key bits from both the bases and obtain the standard BB84 result as a particular case when $p_X=p_Z$. This adaptability is useful in practice, as the optimal ratio between the bases can depend on the characteristics of the quantum channel.

In a single key session, N pulses are sent by the transmitter to the receiver (Bob). Because of channel and detector losses, only $C \leq N$ non-empty counts are registered by Bob in each session. From these counts, the users distil the final key, through a series of classical procedures which require communication over a public channel. These include *sifting*, to select the non-empty counts with matching bases; *error correction* (EC), to determine the number of transmission errors, E, in the non-empty counts and correct them; *privacy amplification* (PA), to remove from a potential eavesdropper (Eve) the information which has possibly leaked to her; *authentication* and *verification*, to prevent man-in-the-middle attacks and guarantee that the users strings match with probability arbitrarily close to 1.

All this information can be used to split the numbers N, C and E, into smaller groups, according to the users choices of the basis and of the intensity for each pulse, thus realising the *advanced data analysis* necessary for the T12 protocol. Whenever the bases do not match, the data are discarded through the sifting procedure. From the results with matching bases, the quantities summarized in Table 1 can be drawn.

Table 1. Quantities for the advanced data analysis of the T12 protocol

| | |
|---|---|
| $N_{\mu_j ZZ}, N_{\mu_j XX}$ | Nr. of pulses with intensity $\mu_j$ and basis Z, X |
| $C_{\mu_j ZZ}, C_{\mu_j XX}$ | Nr. of non-empty counts from pulses with intensity $\mu_j$ and basis Z, X |
| $E_{\mu_j ZZ}, E_{\mu_j XX}$ | Nr. of errors in the non-empty counts from pulses with intensity $\mu_j$ and basis Z, X |

The number of pulses N satisfies the following relations: $N = \sum_{j=\{0,1,2\}} N_{\mu_j}$, $N_{\mu_j} = N_{\mu_j ZZ} + N_{\mu_j ZX} + N_{\mu_j XZ} + N_{\mu_j XX}$. Similar relations hold for C and E. These quantities and the knowledge of $\mu_j$ are used to assess the security of the protocol in the finite-size scenario. The final rate of the protocol is given by the sum of the two secure key rates distilled from the signal pulses separately in the two bases.

## 3. Secure key rate

In this section we determine the rate equation for the T12 protocol using the proof method of [16], later extended in [26]. The proof starts by providing an entanglement-based description of the preparation and distribution stage of the T12 protocol.

The N pulses of the T12 protocol are prepared by Alice using an attenuated laser that emits a series of weak coherent states with random phases. She also varies the intensity of the laser in order to realise the decoy-state technique. This source has been shown to be statistically equivalent to the preparation of N entangled states $\rho_{A^N B^N}$ followed by Alice's measurement of the subspace A [35]. Such a measurement can be done at any time, so it is possible to postpone it at the very end of the protocol without loss of generality, thus remaining with an entangled state shared by the users. The single entangled state $\rho_{AB}$ contributing to $\rho_{A^N B^N}$ can be written as [35] $\rho_{AB} = |\Phi_D^{(k)}\rangle_{AB} \langle \Phi_D^{(k)}|$, where D={Z, X} is a basis index, k is the number of photons in the light pulses, and $|\Phi_D^{(k)}\rangle_{AB} = (|0_D\rangle_A |0_D^{(k)}\rangle_B + |1_D\rangle_A |1_D^{(k)}\rangle_B)/\sqrt{2}$. The states $|0_D\rangle$ and $|1_D\rangle$ have been defined in the previous section and the states $|0_D^{(k)}\rangle$ and $|1_D^{(k)}\rangle$ are k-photon number states in the basis D with bit values 0 and 1, respectively. As a result, the entanglement based description adopted in [16] holds for the T12 protocol.

As a second step, we assume that Bob's detectors are threshold detectors with equal efficiency. When no detector clicks, an empty count is registered by Bob, while all the other cases are non-empty counts. This can be treated as a binary positive-operator valued measure (POVM), a particular 2-dimensional case of the proof method in [16]. Also, the results from Bob's measurement are non-ambiguous if we assign orthogonal outcomes to the two detectors and double counts to one of the two detectors, chosen at random. Overall, this description of the T12 protocol detection stage coincides with that adopted in various security proofs, among which those more relevant to the present paper, described in [16, 17, 26, 35].

After $N$ signals have been distributed, $C$ non-empty counts are detected by Bob. Using a public (authenticated) channel, the users run the sifting procedure, in which they discard the data corresponding to empty counts and non-matching bases and remain with a pair of raw keys. By performing the PE procedure, they can compute from their data the statistics $\lambda_{(a,b)}$, i.e. the frequencies of the detected symbols and of the errors in the detected symbols. This will let them infer the maximum information gained by Eve during the key session. After that, EC and PA will complete the key distillation procedure and provide them with the final key.

In [16], it is shown how to calculate in the finite-size scenario, under the assumption of collective attacks, the *rate per detected qubit* $r = L/n$ for any protocol which comply with the description given above. This legitimates us to use this method for estimating the T12 protocol secure key rate in the finite-size case. In the definition of $r$, the symbol $L$ is the number of secure bits after PA, while $n$ is the number of raw key bits before EC that will contribute to the final key. In the T12 protocol, it is $n = C_{uZZ}$ or $n = C_{uXX}$ according to whether the secure bits are distilled from the Z basis or from the X basis, respectively.

In what follows, we assume for simplicity that the secure bits are distilled from the Z basis. Therefore we omit the basis label. However, all the results hold for the X basis too and the final rate is given by the sum of the two rates from the two separate bases. From [16], we can write the rate per detected qubit as follows:

$$r = H_{\xi_{PE}}(A|E) - (\text{leak}_{EC} + \Delta)/n, \qquad (1)$$

with

$$H_{\xi_{PE}}(A|E) = \min_{\sigma_{AE} \in \Gamma_{\xi_{PE}}} H(A|E), \qquad (2)$$

and

$$\Delta = 7\sqrt{n \log_2\left(\frac{2}{\varepsilon_s - \varepsilon_{PE}}\right)} + 2\log_2\left[\frac{1}{2(\varepsilon - \varepsilon_s - \varepsilon_{EC})}\right]. \qquad (3)$$

The term $\text{leak}_{EC}$ in Eq. (1) accounts for the number of bits publicly transmitted during EC; it is a classical quantity and can be directly measured in the experiment. A common way to appraise it is by using the expression $\text{leak}_{EC} = nf_{EC}h(Q_Z)$, where h(.) is the binary entropy and $Q_Z$ is the bit error rate measured in the Z basis. The parameter $f_{EC} \geq 1$ accounts for the EC efficiency. In Eq. (2), the conditional von Neumann entropy $H(A|E)$ represents Eve's uncertainty about Alice's string. It has to be minimised over all possible Alice-Eve joint states $\sigma_{AE}$ which are contained in a set $\Gamma_{\xi_{PE}}$, specified later on together with the other quantities appearing in Eq. (3). The entropy $H(A|E)$ has been explicitly given in equation (13) of [26] under the same protocol description given so far. After translating that result into our notation, it reads:

$$H(A|E) = \tilde{g}_{uZ}^{(0)} + \tilde{g}_{uZ}^{(1)}[1 - h(\tilde{q}_X^{(1)})], \qquad (4)$$

where

$$\tilde{g}_{uZ}^{(k)} = \frac{\tilde{y}_Z^{(k)} \cdot (e^{-u} u^k / k!)}{(C_{uZZ} / N_{uZZ})}. \qquad (5)$$

In Eq. (4), $\tilde{q}_X^{(1)}$ is the error rate in the X basis of the pulses containing 1 photon. In Eq. (5), the quantity $\tilde{y}_Z^{(k)}$ is the k-photon yield, i.e. the conditional probability that Bob registers a count when Alice emits k photons, in the Z basis. These quantities are indicated with a tilde to recall that they have to be optimised in the finite-size setting in order to obtain the final key rate, as prescribed by the minimisation procedure contained in Eq. (2).

It is worthy to point out that the result in Eq. (4) has been initially obtained by Koashi [35], under the same entanglement-based description of an efficient BB84 protocol with decoy states and imperfect devices as the one adopted thus far for the T12 protocol. Specifically, the Heisenberg uncertainty principle was used to show that Eve's uncertainty about Alice's string distilled in the Z basis can be quantified as $1 - H_X$, with $H_X = 1 - g_Z^{(0)} - g_Z^{(1)}[1 - h(q_X^{(1)})]$ (see Eq. (9) of [35]). In fact, this leads to an expression equal to that in Eq. (4).

The T12 protocol key rate, i.e. the amount of secure information per qubit, in the finite-size scenario, for the Z basis, can then be obtained by replacing Eqs. (2) – (5) and the explicit expression for $\text{leak}_{EC}$ into Eq. (1), and multiplying $r$ by the detection rate of the signal pulses in the Z basis:

$$R_Z = (C_{uZZ} / N_{uZZ}) \cdot r$$

$$= \min\{e^{-u} \tilde{y}_Z^{(0)} + ue^{-u} \tilde{y}_Z^{(1)}[1 - h(\tilde{q}_X^{(1)})]\} - \frac{C_{uZZ}}{N_{uZZ}} f_{EC} h(Q_Z) - \frac{\Delta}{N_{uZZ}} \qquad (6)$$

$$= \{e^{-u} \underline{y}_Z^{(0)} + ue^{-u} \underline{y}_Z^{(1)}[1 - h(\overline{q}_X^{(1)})]\} - \frac{C_{uZZ}}{N_{uZZ}} f_{EC} h(Q_Z) - \frac{\Delta}{N_{uZZ}} \qquad (7)$$

As said, an equation analogous to $R_Z$ can be obtained for $R_X$ by swapping the Z and X labels in all equations. In Eqs. (6), (7), capital letters indicate quantities which are directly measurable in the experiment, while the small letters are for parameters which have to be indirectly estimated. Notice that the minimisation in Eq. (2) has translated into that of Eq. (6), which in turn is accomplished in Eq. (7) by minimising $\tilde{y}_Z^{(0)}$, $\tilde{y}_Z^{(1)}$ and maximising $\tilde{q}_X^{(1)}$ in their variability range according to a worst-case treatment. In Eq. (7) we have introduced the notation $\underline{y}_Z^{(k)} = \min_{I_{\varepsilon_{PE_Y}}}[\tilde{y}_Z^{(k)}]$, k={0,1}, and $\overline{q}_X^{(1)} = \max_{I_{\varepsilon_{PE_B}}}[\tilde{q}_X^{(1)}]$, where $I_{\varepsilon_{PE_Y}}$, $I_{\varepsilon_{PE_B}}$ are confidence intervals for the quantities $\tilde{y}_Z^{(k)}$, k={0,1} and $\tilde{q}_X^{(1)}$, respectively. How to find such intervals and perform the optimisation will be explained in the next section. Here, we tighten up some loose ends of the above discussion.

The conditional von Neumann entropy in Eq. (2) is the result of a lower bound to the smooth min entropy $H_{\min}^{\varepsilon_s}\left(A^n | E^n\right)$ given in Lemma 2 of [16]. Under the assumption of collective attacks by Eve, this bound holds for any protocol which can be described as an entanglement-distribution protocol and adopts the proper PE procedure to compute the statistics $\lambda_{(a,b)}$. We have already shown the entanglement-based description of the T12 protocol. The discussion of the PE procedure involves the definition of the set $\Gamma_{\xi_{PE}}$ in Eq. (2), which, in [16], was defined via the following relation:

$$\Gamma_{\xi_{PE}} = \left\{\sigma_{AE} : \|\lambda_m - \lambda_\infty\| \leq \xi_{PE}\right\}. \tag{8}$$

The meaning of Eq. (8) is that if the statistics $\lambda_m$, acquired from a dataset of size m, is closer than $\xi_{PE}$ to the asymptotic statistics $\lambda_\infty$, obtained from an infinite dataset, then the state $\sigma_{AE}$ belongs to $\Gamma_{\xi_{PE}}$, except with a probability $\varepsilon_{PE}$. In this case, $\sigma_{AE}$ is a legitimate state for the minimisation procedure set forth in Eq. (2) and then in Eq. (6).

In order to turn Eq. (8) into an operative definition, it is necessary to find a relation between the upper bound to the variation distance $\xi_{PE}$ and the probability $\varepsilon_{PE}$ that PE will fail to select the correct set of states for the minimisation of Eq. (2). In [16], this relation was governed by a Lemma through the law of large numbers. We limit here to the case of a binary POVM, because this is the only one present in the T12 protocol. In this case, it was proven that if $\xi_{PE} = \sqrt{[2\ln(m+1/\varepsilon_{PE})]/m}$ then $\Pr(\|\lambda_\infty - \lambda_m\| > \xi_{PE}) \leq \varepsilon_{PE}$. Hence, the subsequent assignment $\lambda_m = \lambda_\infty \pm \xi_{PE}$ was used to select an interval around $\lambda_\infty$ in which the true value of the statistics falls with probability $1 - \varepsilon_{PE}$. Such a method clearly requires the knowledge of the asymptotic statistics $\lambda_\infty$, without which the finite-size statistics $\lambda_m$ cannot be drawn. In [26], $\lambda_\infty$ was taken from the analytical expressions given in [23]. Then the m-size statistics was obtained from the assignment $\lambda_m = \lambda_\infty \pm \xi_{PE}$, as said, with the sign chosen according to whether a lower or an upper bound was needed for the quantities appearing in a rate equation analogous to our Eq. (7).

In our solution, we use a statistical method to directly estimate the confidence interval for $\lambda_\infty$, without the need of knowing $\lambda_\infty$ itself. We call $I_{\varepsilon_{PE}}$ the confidence interval which contains the true value of the statistics with probability $1 - \varepsilon_{PE}$. To determine it, we initially set a confidence level equal to $1 - \varepsilon_{PE}$ and then we use this value to directly solve the optimisation problems of Eq. (7), as explained in the next section. This provides the upper and lower bounds, $\lambda^+$ and $\lambda^-$, of the confidence interval for $\lambda_\infty$, formally, $I_{\varepsilon_{PE}} \equiv [\lambda^-, \lambda^+]$.

For an easy comparison with [16], we note that our approach would be equivalent to choosing $\xi_{PE}^- = \lambda_\infty - \lambda^-$ ($\xi_{PE}^+ = \lambda^+ - \lambda_\infty$) for a parameter that needs to be minimised (maximised) in Eq. (7). In this case, the exact value of $\lambda_\infty$ would not be relevant anymore, because of the assignment $\lambda_m = \lambda_\infty - \xi_{PE}^-$ ($\lambda_m = \lambda_\infty + \xi_{PE}^+$), which removes $\lambda_\infty$ from the problem and leaves $\lambda_m = \lambda^-$ ($\lambda_m = \lambda^+$). From this, it is straightforward to see that $\Pr(\|\lambda_\infty - \lambda_m\| > \xi_{PE}^\pm)$ coincides with the probability that the true value of the statistics falls outside $I_{\varepsilon_{PE}}$. This probability is, by construction of $I_{\varepsilon_{PE}}$, less than or equal to $\varepsilon_{PE}$, as required in [16].

Finally, we should complete the description of the parameters given in Eq. (3). The smoothing parameter $\varepsilon_s \geq 0$ descends from the smooth min entropy $H_{min}^{\varepsilon_s}(A^n | E^n)$, used after Lemma 1 in [16] to quantify the number of uniform bits that can be extracted from an error-corrected key using PA. A perfectly uniform distribution of the output bits is attainable only in the asymptotic limit. In the finite-size case, it is necessary to accept some deviation from the ideal uniform distribution, quantified by $\varepsilon_s$. The failure probability of the PE procedure is indicated as $\varepsilon_{PE}$. It has the meaning mentioned above that the estimated confidence interval $I_{\varepsilon_{PE}}$ fails to contain the true value of the statistics. Generally speaking, the wider the interval, the lower $\varepsilon_{PE}$. Moreover, it is intuitive that smaller values of $\varepsilon_{PE}$ correspond to a higher security level and to a lower key rate. The probability that EC fails is denoted by $\varepsilon_{EC}$. However, it should be said that EC includes also the verification step in which the users verify that their keys are equal. So $\varepsilon_{EC}$ more properly represents the probability that EC fails to correct the key, with the users unaware of this failure. Finally, the total failure probability of the system is $\varepsilon$, which is given by the sum of the other failure probabilities, $\varepsilon = \varepsilon_{EC} + \varepsilon_s + \varepsilon_{PE}$. It amounts to $10^{-10}$ in the present work. We choose and fix the numerical values of all the epsilon values so to fulfil the chain relation $\varepsilon - \varepsilon_{EC} > \varepsilon_s > \varepsilon_{PE} \geq 0$, as required in [16].

## 4. Finite-size statistical analysis and parameter estimation

The T12 protocol rate in Eq. (7) contains three optimisation procedures which have to be performed over a range specified by the statistics acquired during the PE procedure. In this section, we give a more detailed description of such a procedure.

In the asymptotic setting, the decoy state technique allows to put the measurable quantities in one-to-one correspondence with the parameters to be estimated [24]:

$$Y_{\mu Z} = \sum_k \frac{e^{-\mu} \mu^k}{k!} y_Z^{(k)}, \qquad (9)$$

$$B_{\mu X} = \sum_k \frac{e^{-\mu} \mu^k}{k!} y_X^{(k)} q_X^{(k)}. \qquad (10)$$

With the implicit assumption that the single terms in the sums, e.g. $y_Z^{(k)}$, do not depend explicitly on $\mu$ [23], the above set of equations allows to determine all the unknown parameters exactly. The measurable quantities are on the LHS while the unknowns are on the RHS, coupled with the Poissonian distribution of the photon number. $Y_{\mu Z}$ is the rate of Bob's detections when a pulse with average photon number $\mu$ was prepared and the Z basis used. $B_{\mu X}$ is the bit error rate measured from pulses with average photon number $\mu$, in the X basis.

In the finite-size setting, there is no more exact correspondence between measured quantities and estimated parameters and the acquired statistics allows for different realisations of $y_Z^{(k)}$ and

$q_X^{(k)}$. Among these, security imposes to select the one which maximises Eve's information through a constrained optimisation process.

A first optimisation is required because there is only a finite number of intensities that Alice can prepare. A second optimisation is required because the data sample itself is finite. These two processes are usually carried out separately: the former gives rise to analytical estimates [23-25] which are then corrected to take the latter into account [26, 29]. While this can provide positive key rates, it might not be the ideal strategy. In fact, there is no guarantee that the analytical estimates are optimal. Moreover, the numerical fluctuations in the finite sample are often treated using an approximate method, like the Wald statistical test [36], in which the actual coverage probability of the confidence interval is much smaller than expected [37].

To overcome these limitations, we use a combination of statistical analysis and constrained optimisation [38, 39]. The statistical analysis relies on the Clopper-Pearson (CP) test [40], which avoids approximations and provides the most conservative confidence interval compatible with the statistics acquired from the PE procedure. This allows us to map the experimental quantities of Table 1 into useful bounds with confidence level $1-\varepsilon_{PE}$. Then, using these bounds as constraints, we can obtain the estimated quantities through an optimisation algorithm. Both these steps can be efficiently performed via standard numerical routines. In Table 2 we summarise the relevant steps and quantities involved in this approach.

**Table 2. Relations among experimental, statistical and estimated quantities.**

| LAYER | PROCESS | QUANTITIES |
|---|---|---|
| Experiment | Measurements | $N_{\mu_j ZZ}, N_{\mu_j XX}, C_{\mu_j ZZ}, C_{\mu_j XX}, E_{\mu_j ZZ}, E_{\mu_j XX}$ |
| Statistics | Clopper-Pearson with confidence $1-\varepsilon_{PE}$ | $Y^-_{\mu_j Z}, Y^+_{\mu_j Z}, B^-_{\mu_j X}, B^+_{\mu_j X}$ |
| Estimation | Constrained optimisation | $\underline{y}^{(0)}_Z, \underline{y}^{(1)}_Z, \overline{q}^{(1)}_X$ |

When the data sample is finite, Eqs. (9), (10) turn into a set of inequalities which can still be used to guarantee the security of the protocol [25, 38, 39]. In fact, they can be rewritten for the three intensity levels $\mu_j$ of the T12 protocol as:

$$Y^-_{\mu_j Z} \leq \sum_k \frac{e^{-\mu_j}(\mu_j)^k}{k!} \tilde{y}^{(k)}_Z \leq Y^+_{\mu_j Z}, \quad j=\{0,1,2\}, \quad (11)$$

$$B^-_{\mu_j X} \leq \sum_k \frac{e^{-\mu_j}(\mu_j)^k}{k!} \tilde{y}^{(k)}_X \tilde{q}^{(k)}_X \leq B^+_{\mu_j X}, \quad j=\{0,1,2\}. \quad (12)$$

The quantity $Y^\pm_{\mu_j Z}$ ($B^\pm_{\mu_j X}$) contains lower and upper bounds to the detection rate (error rate) of the pulses with intensity $\mu_j$ measured by the users in the Z (X) basis. From this, we can draw the confidence intervals introduced in Section 3, $I_{\varepsilon_{PE_Y}} = \left[Y^-_{\mu_j Z}, Y^+_{\mu_j Z}\right]$ for the detection rate and

$I_{\varepsilon_{PE_B}} = \left[ B^-_{\mu_j X}, B^+_{\mu_j X} \right]$ for the error rate, each with the appropriate confidence level coming from the CP test.

It is easy to see that under the assumption of independent and identically distributed (i.i.d) variables, the quantities measured in our QKD experiment follow the Binomial distribution. In fact, to measure, e.g., the transmission errors, a binary POVM is employed which determines whether the users share the same bit (success) or not (failure). The same is true for the output of Bob's threshold detector which, as said in Section 3, can only give as a response either "click" or "no-click". As a result, a series of Bernoulli trials is obtained from the experiment and the bounds can be deduced by the CP method for all the intensity values and for the Z basis (the X basis is analogous) via the following equations:

$$Y^-_{\mu_j Z} = \beta\left( \frac{\varepsilon_{PE_Y}}{2}; C_{\mu_j ZZ}, N_{\mu_j ZZ} - C_{\mu_j ZZ} + 1 \right), \quad j = \{0,1,2\}, \tag{13}$$

$$Y^+_{\mu_j Z} = \beta\left( 1 - \frac{\varepsilon_{PE_Y}}{2}; C_{\mu_j ZZ} + 1, N_{\mu_j ZZ} - C_{\mu_j ZZ} \right), \quad j = \{0,1,2\}. \tag{14}$$

Here $\beta(\alpha; s, t)$ is the α-th quantile from a beta distribution with shape parameters $s$ and $t$, which are related to the measurable quantities of Table 1. The bounds $B^-_{\mu_j X}$, $B^+_{\mu_j X}$ are found in a similar way, using $N_{\mu_j XX}$, $E_{\mu_j XX}$ and $\varepsilon_{PE_B}$.

Now we can use constrained optimisation to minimise $\tilde{y}^{(0)}_Z$, $\tilde{y}^{(1)}_Z$ and maximise $\tilde{q}^{(1)}_X$ in their respective confidence intervals. The explicit problems are shown in the Appendix. The solutions have already been introduced in Section 3 and written in Eq. (7) as $\underline{y}^{(k)}_Z$, k={0,1}, and $\overline{q}^{(1)}_X$. Because objective functions and constraints in the optimisation problems are all reduced to linear functions, the solutions are guaranteed to be global maxima or minima [41]. This ensures that when they are finally plugged into Eq. (7), the obtained finite-size secure key rate of the T12 protocol is an absolute minimum, given the experimental statistics acquired by the users, i.e. a worst-case bound to the real rate.

## 5. Experimental implementation and numerical simulation

In this section, we describe the experimental implementation of the T12 protocol in a high bit rate QKD system. In all the experiments, we set $\varepsilon = 10^{-10}$. Moreover, we choose the probability of the minority basis to be $p_X = 2^{-4} = 1/16$. To determine this value, we initially run a simulation and find the value $p_X^{opt}$ that maximises the secure key rate. Then, for experimental convenience, we choose the power-of-2 value which is the closest to $p_X^{opt}$. Following the same logic, we choose the intensity probabilities as $p_w = 2^{-8} = 1/256$, $p_v = 2^{-7} = 1/128$ and $p_u = 1 - p_w - p_v$. The values for the average photon number are $u = 0.425$, $v = 0.044$, $w = 0.001$. They are derived from a simulation under the constraint of being compatible with the actual intensity modulator used in the setup. Small variations of these values do not affect the overall secure rate, thus showing that they are all close to optimal.

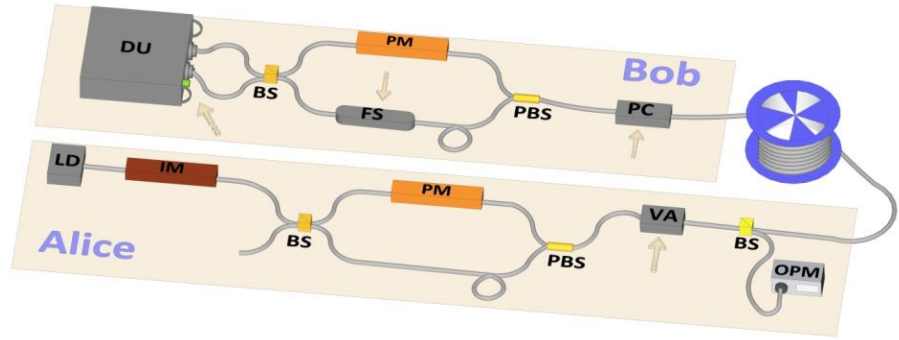

Fig. 1. Experimental setup for the T12 protocol. In Alice's layout, light pulses are emitted by a 1550 nm laser diode (LD), pulsed at 1 GHz, and transmitted through an intensity modulator (IM) and an unbalanced Mach-Zehnder interferometer. This is composed by a fibre-integrated beam-splitter (BS), a phase modulator (PM) and a final polarising BS (PBS). A variable attenuator (VA) is used to set the intensity of the pulses at the desired level. An optical power meter (OPM) measures the total flux in the fibre and adjust the VA in real-time in order to keep it constant. After a fibre spool of different lengths, the light passes through a polarization control (PC) and a second interferometer that matches Alice's. In one arm, a fibre-stretcher (FS) is used to match the arms length between the two distant interferometers, thus generating interference at the final BS. Pulses are eventually measured by a detection unit (DU).

The QKD experimental setup is shown in Fig. 1. The system operates in the telecom band at a wavelength of 1550 nm. On Alice's side, encoding is accomplished through phase modulator voltages corresponding to $(0, \pi)$ for the Z basis and $(\pi/2, 3\pi/2)$ for the X basis. Bob uses a $(0, \pi/2)$ modulation for decoding. The two communicating parties are linked together via a dispersion shifted fibre which has a dispersion coefficient of 4 ps/(km·nm). Single photons are detected by gated InGaAs avalanche photodiode (APD) in the self-differencing mode [42]. The APDs are cooled thermoelectrically to –30°C and operated at a detection efficiency of 20.5%, dark count probability per gate of $2.1\times10^{-5}$ and after pulse probability 5.25%. A software program controls all the equipment continuously, calculates the QBER for the fibre-stretcher to counteract any drift in the phase and corrects the detector gate delay and polarisation so as to maximise the count rate. The average photon number is stabilised using the feedback from a power meter connected to the main channel through a beam splitter with fixed known splitting ratio.

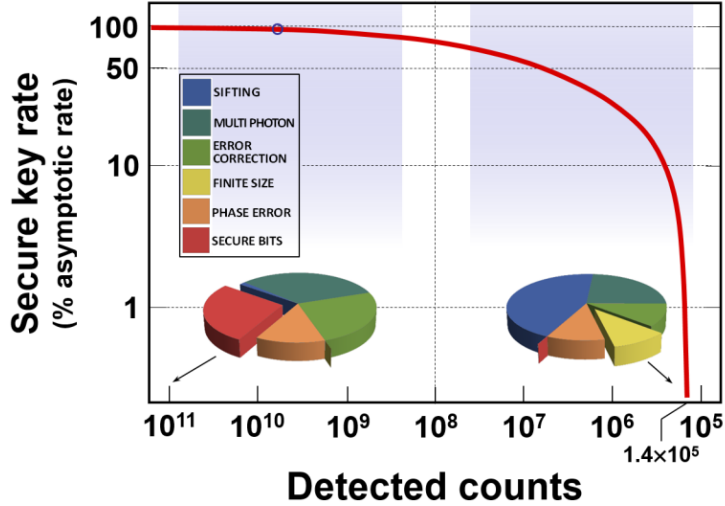

Fig. 2. Secure key rate of the T12 protocol versus the size of the detected data sample, for $\varepsilon = 10^{-10}$. Parameters are obtained with the setup of Fig. 1 from an experimental QKD run over a 50 Km optical fibre. The key rate, depicted with a red line, is measured in percentage from the asymptotic value. Pie-charts with the breakdown of the counts leading to the final secure bits are given for sample sizes of $10^{11}$ counts and $1.4 \times 10^5$ counts. The latter value is the smallest sample size providing a positive rate and can be acquired in less than 60 ms with our system. For a typical acquisition time of 20 minutes, the block size is $\sim 5 \times 10^9$ counts and the corresponding secure key rate is indicated by a violet circle on the red line. In addition, the minority basis probability has been optimised so to obtain the highest secure key rate at all distances. It goes from $p_X = 0.0065$ for a sample of $10^{11}$ counts to $p_X = 0.2290$ for a sample of $1.4 \times 10^5$ counts.

The experimental parameters given above have been used to numerically simulate the secure key rate of the T12 protocol as a function of the detected sample size on a fibre distance of 50 Km. This is shown in Fig. 2, with the sample size decreasing from left to right. The secure key rate is calculated by applying Eq. (7), separately to each basis, and then adding up the two resulting partial rates. As expected, it decreases when the sample size becomes smaller. However, already for a sample size of $10^8$ counts, the protocol performs at about 85% of its asymptotic value. Moreover, the key rate remains positive up to a sample size of $1.4 \times 10^5$ counts. With the same experimental parameters, the security proof by Cai and Scarani [26] would provide 50% of the asymptotic rate for a sample size of $10^8$ counts, and $1.6 \times 10^7$ counts as the minimum size tolerated by the protocol. Therefore our model is significantly more resilient to finite size effects. In Fig. 2, we also show two pie-charts with the breakdown of the counts collected in the experiment, one for a large data sample ($10^{11}$ counts) and the other for the minimum data sample tolerated by the protocol ($1.4 \times 10^5$ counts). The total detected sample is reduced by basis sifting, EC and PA applied to multi-photon events and phase-errors before reaching the final secure key fraction. The detrimental contribution of the finite-size effects becomes apparent for smaller sample sizes.

## 6. Experimental results

With the setup shown in Fig. 1, a series of experiments were conducted to verify the feasibility and practicality of the T12 protocol. The main feature to be demonstrated is the capability to measure the quantities in Table 1 which enter the secure key rate of the protocol. This requires an *advanced* data analysis of the experimental sample, which takes into account not only the basis information [21, 44], but also the intensity information. To our knowledge, such an advanced data analysis has never been shown before in QKD.

In Fig. 3, we report the sifted count rates for the two bases Z and X, for the three intensity values $u$, $v$ and $w$. In the lower part of the figure, we show the QBERs in the two bases, for the intensity $u$. The error rates related to $v$ and $w$ do not enter the secure key rate of the protocol (see Appendix) and are not displayed. Data are measured over an optical fibre length of 50 km, with key sessions of 20 minutes and for a total time of more than 3 hours. In this period, the average sifted counts, expressed in counts per session, are $C_{uZZ} = (5.016 \pm 0.082) \times 10^9$, $C_{uXX} = (2.231 \pm 0.035) \times 10^7$, $C_{vZZ} = (6.21 \pm 0.11) \times 10^6$, $C_{vXX} = (2.843 \pm 0.051) \times 10^4$, $C_{wZZ} = (1.259 \pm 0.017) \times 10^6$, $C_{wXX} = (5.79 \pm 0.10) \times 10^3$ while the average error rates are $Q_{uZZ} = E_{uZZ}/C_{uZZ} = (4.26 \pm 0.20)\%$ and $Q_{uXX} = E_{uXX}/C_{uXX} = (3.64 \pm 0.65)\%$. The mean value of the error rates is determined mainly by the modulation errors, lowest in the X basis, while the standard deviation is smaller in the Z basis, due to the larger sample from which the statistics is drawn. From Fig. 3 we can also see that the sample collected in a 20 minute session is of about $5 \times 10^9$ counts, therefore implying from Fig. 2 that the reduction due to finite-size effects is negligible.

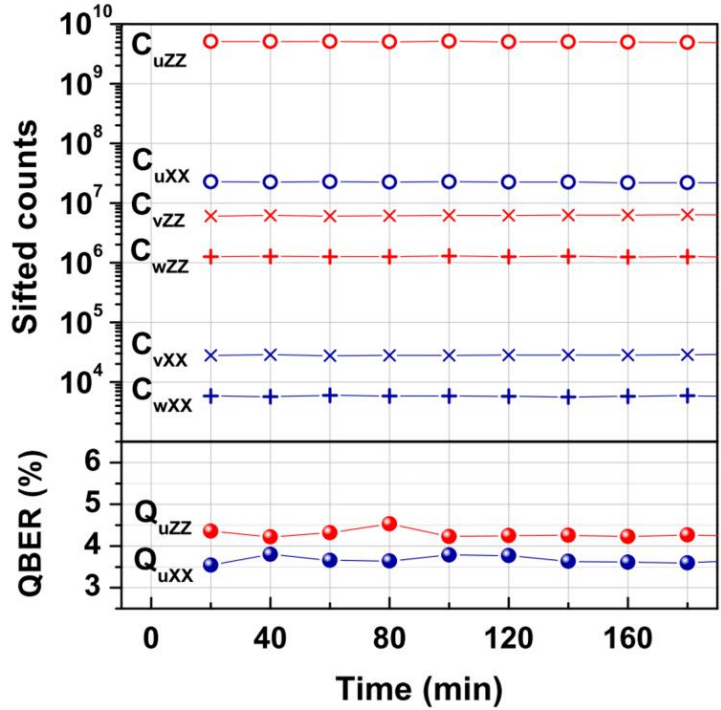

Fig. 3. Advanced data analysis for the T12 protocol. Displayed are the basis-dependent and intensity-dependent counts and QBERs acquired by the system. The quantum channel is a 50 km optical fibre and the key session time is 20 minutes.

From the measured quantities, the secure key rate of the T12 protocol is automatically calculated in the system. We display it in Fig. 4, under the same experimental conditions as in Fig. 3. For comparison, we also show the key rate of the standard BB84 protocol, implemented by changing the basis bias to $p_X = p_Z = 1/2$. Notice that to calculate the key rate of the BB84 protocol, we can use exactly the same rate equation as for the T12 protocol. For both protocols the total key rate is given by the sum of the key rates distilled separately in the two bases. Hence we also show in Fig. 4 the basis-dependent key rates. The total rate for the T12 protocol is 1.09 Mbps and is distilled almost entirely from the majority basis Z, while the total rate for BB84 is 0.63 Mbps and the two bases contribute nearly equally to form it.

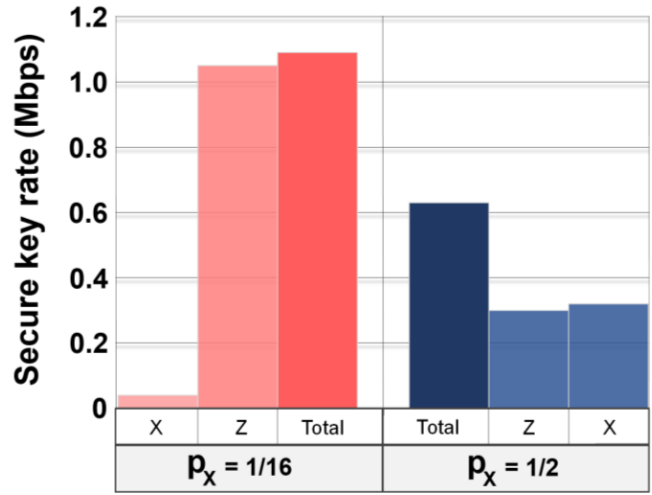

Fig. 4. Experimental total and basis-dependent secure key rates. For the T12 protocol ($p_X$=1/16) data are in red colours. For the standard BB84 protocol ($p_X$=1/2) data are in blue colours. Experimental data are obtained over a 50 km optical fibre link and a typical acquisition time of 20 minutes. Each of the total key rates is given by the sum of the respective basis-dependent key rates.

Because in the T12 protocol $p_X = 1/16$, only 11.7% of the detected counts are discarded, as opposed to 50% in the BB84 protocol. This increases the theoretical efficiency of the T12 protocol to 88.3%, which is 76.6% higher than the standard BB84 protocol. The experimental results in Fig. 4 agree well with this value. In fact, the obtained enhancement is 73.5%, very close to the theoretical limit.

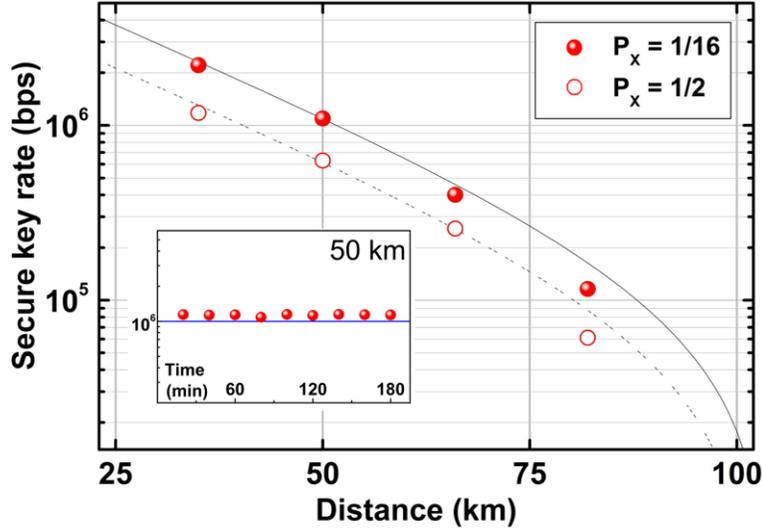

Fig. 5, Experimental secure key rates as a function of fibre distance. The fibre distances experimentally tested are 35, 50, 65 and 80 km, respectively. For the T12 protocol ($p_X$=1/16, filled circles) the values of the secure key rate at these distances are 2.20, 1.09, 0.40 and 0.12 Mbps. For the standard BB84 protocol ($p_X$=1/2, empty circles) they are 1.18, 0.63, 0.26 and 0.06 Mbps. The solid and dashed lines are theoretical curves for T12 and BB84 protocol, respectively. The inset shows the real-time secure key rate provided by the T12 protocol at 50 km, over more than 3 hours of continuous operation. All the samples at this distance feature secure key rates exceeding 1 Mbps.

Finally, in Fig. 5, we show the total secure key rate as a function of the fibre distance, for both the T12 protocol and the BB84 protocol. Data was acquired using 4 different lengths of fibre spool connecting the users, 35, 50, 65 and 80 km, respectively. For a fair comparison at all distances, we kept the acquisition time per data block fixed, equal to 20 minutes. The T12 secure key rates are, for increasing distances, 2.20, 1.09, 0.40 and 0.12 Mbps, the highest reported to date in QKD literature. With respect to previous experiments with finite-size security on BB84-like protocols [44-46], these results represent an improvement of 3 orders of magnitude [46] or more [44, 45].

## 7. Conclusion

We have investigated the finite-size security of the efficient version of the BB84 protocol, implemented with attenuated laser and decoy state technique. For that, we have purposely defined a protocol, namely the T12, which includes several practical advantages.

After gathering all aspects under a unified description, we have applied the Scarani and Renner proof method [16] to assess the protocol secure key rate. Then we have adopted a more efficient parameter estimation procedure to improve the resistance of the protocol to finite-size effects. As a result, the minimum data sample size providing a positive key rate has now greatly improved, by more than two orders of magnitude. Numerical simulations have been provided, showing explicitly the effect of the data sample finiteness on the secure key rate. With samples of $10^8$ bits, finite-size effects reduce the asymptotic key rate by about 15%. For higher sample sizes, the reduction becomes negligible.

We have then implemented the T12 protocol using a gigahertz-clocked QKD system. All the main features of the protocol, including the high key rate and the increased efficiency over the standard BB84 protocol, have been experimentally tested and agree well with the theoretical predictions.

To demonstrate the correct execution of the protocol, we have shown the results from an advanced data analysis, in which experimental quantities are acquired and processed in real time according to their basis and intensity information.

The high bit rate of the system allows to collect $5\times10^9$ counts in a typical session of 20 minutes thus providing a key rate that is not appreciably affected by the finiteness of the sample. Despite its large security level, represented by a failure probability $\varepsilon=10^{-10}$, the system provides the highest secure key rates reported to date over tens of kilometres in optical fibre.

**Acknowledgments**

Discussions with R. Renner, M. Tomamichel, N. Lütkenhaus, S. Kunz-Jacques are gratefully acknowledged. The work is partly supported by the Commissioned Research of National Institute of Information and Communications Technology (NICT), Japan. K. A. Patel acknowledges personal support via the EPSRC funded CDT in Photonics System Development.

**Appendix**

Here we give the explicit optimisation problems solved by the automatic numerical routine in our system (see also [27, 38]). The optimisation problem for the quantities $\tilde{y}_Z^{(0)}$, $\tilde{y}_Z^{(1)}$ and $\tilde{q}_X^{(1)}$ can be easily written when the bounds from the CP approach [see e.g. Eqs. (13) and (14)] become available. We write the problem for $\tilde{y}_Z^{(0)}$ as:

$$\text{minimise:} \quad \tilde{y}_Z^{(0)} \tag{15}$$

$$\text{subject to:} \quad 0 \leq \tilde{y}_Z^{(0)}, \tilde{y}_Z^{(1)}, ..., \tilde{y}_Z^{(k)} \leq 1, \tag{16}$$

$$Y_{\mu_j Z}^- \leq \sum_k \frac{e^{-\mu_j}(\mu_j)^k}{k!}\tilde{y}_Z^{(k)} \leq Y_{\mu_j Z}^+, \; j=\{0,1,2\}. \tag{17}$$

This constrained optimisation problem is linear and can be efficiently solved by a numerical routine based on linear programming. The problem for minimising $\tilde{y}_Z^{(1)}$ is analogous and we do not need to write it explicitly. The maximisation of $\tilde{q}_X^{(1)}$ is in the following problem:

$$\text{maximise:} \quad \tilde{q}_X^{(1)} \tag{18}$$

$$\text{subject to:} \quad 0 \leq \tilde{y}_X^{(0)}, \tilde{y}_X^{(1)}, ..., \tilde{y}_X^{(k)} \leq 1, \tag{19}$$

$$0 \leq \tilde{q}_X^{(0)}, \tilde{q}_X^{(2)}, ..., \tilde{q}_X^{(k)} \leq 1, \; 0 \leq \tilde{q}_X^{(1)} \leq 1/2, \tag{20}$$

$$Y_{\mu_j X}^- \leq \sum_k \frac{e^{-\mu_j}(\mu_j)^k}{k!}\tilde{y}_X^{(k)} \leq Y_{\mu_j X}^+, \; j=\{0,1,2\}, \tag{21}$$

$$B_{\mu_j X}^- \leq \sum_k \frac{e^{-\mu_j}(\mu_j)^k}{k!}\tilde{y}_X^{(k)}\tilde{q}_X^{(k)} \leq B_{\mu_j X}^+, \; j=\{0,1,2\}. \tag{22}$$

The above problem, unlike the previous one, is not linear anymore because quadratic terms are present in the constraints. Hence, it cannot be solved as efficiently as the other one. This can represent a hindrance from an implementation viewpoint. To overcome it, we adopt a worst-case estimation, demonstrated in [27], using only the intensity value pertaining to the signal pulse, $\mu_0 = u$:

$$\tilde{q}_X^{(1)} \leq \bar{q}_X^{(1)} = \frac{B_{uX}^+ - \frac{1}{2}\underline{y}_X^{(0)}}{e^{-u} u \, \underline{y}_X^{(1)}}. \tag{23}$$

Eq. (23) is obtained starting from the original problem in Eqs. (18) − (22) and step-by-step loosening the constraints, so to have at each step a solution which includes (so it is looser than) the previous one. This can only increase the security of the system. We used $\tilde{q}_X^{(0)} = 1/2$, a standard assumption in QKD descending from the fact that the error rate caused by a 0-photon pulse is 50%. The quantities $\underline{y}_X^{(0)}$ and $\underline{y}_X^{(1)}$ are the solutions to minimisation problems analogous to those solved in the Z basis for $\tilde{y}_Z^{(0)}$ and $\tilde{y}_Z^{(1)}$.

---